\documentclass[aps, pra, twocolumn, superscriptaddress, amsmath,  tightenlines, longbibliography]{revtex4-2}

\usepackage{dcolumn}
\usepackage{graphicx}
\usepackage{epstopdf}
\usepackage{mathrsfs}
\usepackage{subfigure}
\usepackage{booktabs}
\usepackage{amsmath,amsfonts}
\usepackage{physics}
\usepackage{dsfont}
\usepackage{amstext}
\usepackage{amssymb}
\usepackage{amsbsy}
\usepackage{bbm}
\usepackage{amsthm}
\usepackage{graphicx}
\usepackage{xcolor}
\usepackage{CJK}
\usepackage[colorlinks,urlcolor=blue,linkcolor=blue,citecolor=blue]{hyperref}
\usepackage{soul}

\usepackage{url}
\usepackage[colorlinks]{hyperref}
\hypersetup{%
	plainpages=true,
	breaklinks=true,       
	hypertexnames=false,  
	pageanchor=true,
	colorlinks=true,
	linkcolor={blue},
	citecolor={blue},
	urlcolor={blue},
	anchorcolor={black}
}

 \makeatletter

\newcommand{\Rmnum}[1]{\expandafter\@slowromancap\romannumeral #1@}
\makeatother

\begin{document}

\title{Quantum batteries in coherent Ising machine}

\author{Jin-Tian Zhang}
\address{School of Physics and Astronomy, Beijing Normal University, and Key Laboratory of Multiscale Spin Physics (Beijing Normal University), Ministry of Education, Beijing 100875, China}

\author{Shuang-Quan Ma}
\address{School of Physics and Astronomy, Beijing Normal University, and Key Laboratory of Multiscale Spin Physics (Beijing Normal University), Ministry of Education, Beijing 100875, China}

\author{Jing-Yi-Ran Jin}
\address{School of Physics and Astronomy, Beijing Normal University, and Key Laboratory of Multiscale Spin Physics (Beijing Normal University), Ministry of Education, Beijing 100875, China}

\author{Tao Liu}
\address{School of Physics and Optoelectronics, South China University of Technology, Guangzhou 510640, China}

\author{Qing Ai}
\email{aiqing@bnu.edu.cn}
\address{School of Physics and Astronomy, Beijing Normal University, and Key Laboratory of Multiscale Spin Physics (Beijing Normal University), Ministry of Education, Beijing 100875, China}

\begin{abstract}
{With intensive studies of quantum thermodynamics, quantum batteries (QBs) have been
proposed to store and transfer energy via quantum effects. Despite many theoretical models,
decoherence remains a severe challenge and practical platforms are still rare. Here, we propose a QB based on the coherent Ising machine, in which the signal field acts as the core energy-storage unit. To clarify the role of quantum coherence in resisting dissipation, we decompose the ergotropy, i.e., the maximum extractable work from the QB, into its coherent and incoherent components. We find that the coherent part decays at a rate roughly half that of the incoherent part, exhibiting much stronger robustness against decoherence. More importantly, the coherent ergotropy and the average charging power reach their respective maxima at essentially the same moment, which defines the optimal instant to switch off the pump field. Finally, by coupling the QB to a two-level system as the load, we demonstrate an efficient energy discharge process of the proposed QB. Our work establishes a realistic and immediately-implementable QB architecture on a mature optical platform, laying a foundation for experimental exploration of quantum energy storage.}
\end{abstract}
\maketitle

\section{Introduction}

As a widely-used energy-storage device, the traditional battery stores and releases energy through electrochemical reactions that involve ion transport between electrodes \cite{vincent1997modern,dell2001understanding}. From the small dry cells in flashlights to the large batteries in electric cars \cite{scrosati2015advances}, batteries are widely used in many aspects of everyday life.
In recent years, the rapid development of quantum technologies, including quantum computing \cite{Lloyd1955PRL,Knill1998PRL,Knill2000PRL}, quantum metrology and sensing  \cite{Giovannetti2006PRL,Blatt2008PRL,Deutsch1983PRL}, has prompted us to ask whether we can exploit quantum coherence and entanglement to realize energy storage and transfer. Inspired by this consideration, the concept of quantum batteries (QBs) was proposed in 2013 \cite{Alicki2013PRE}. QBs use quantum entanglement, and quantum coherence, and other quantum effects to optimize energy storage and transfer \cite{Campaioli2017PRL,Ferraro2018PRL,Andolina2019PRL,Gar2020PRL,Rossini2020PRL,Yang2023PRL,Song2024PRL,Monsel2020PRL,Hovhannisyan2013PRL,Crescente2020PRB,Guryanova2016NC,Barra2019PRL,
Seah2021PRL,Ahmadi2024PRL,Bhattacharyya2024PRL,Campaioli2024RMP,Reinforcement2024PRL,Ghosh2020PRA,Huangfu2021PRE,Gyhm2022PRL,Andolina2019PRB,Farina2019PRB,Salvia2023PRR,Chen2022PRE,Cakmak2020PRE,Santos2023PRA}.

Recently, the design of QB based on different models and the development of high performance QB have attracted broad interest. Various interesting models have emerged, such as collective-spin QB based on the Dicke model \cite{Ferraro2018PRL,Crescente2020PRB}, Heisenberg spin-chain architectures \cite{Ali2024PRA,Ghosh2020PRA,Ghosh2022PRA}, Jaynes–Cummings interaction models \cite{Yang2024PRA}, and coupled resonators \cite{Ma2024PRA,Lu2025PRL}. These studies have found that, under ideal conditions, QB can achieve collective charging, resulting in significantly enhanced charging power and improved efficiency compared to independent charging \cite{Ferraro2018PRL,Yang2023PRL,Kamin2020PRE,Binder2015NJP}. However, the ubiquitous decoherence and environmental dissipation in practical systems inevitably degrade the performance of QB. Therefore, in order to put their application into practice, it is a key issue to develop a platform that simultaneously offers long coherence times and experimental feasibility \cite{Uzdin2016Entropy,Barra2019PRL,Santos2021PRE,Francica2020PRL,Tirone2023PRL}.

To address this issue, we propose a QB based on the coherent Ising machine (CIM) \cite{Wang2013PRA,Marandi2014Nature,Calvanese2021PRL,Inagaki2016Nature}. The CIM solves the ground state of the Ising model, and it has been widely used with the Hamiltonian given by $H_{\rm Ising}=-\sum_{1\leqslant j<l\leqslant N}J_{jl}\sigma_{j}\sigma_{l}$. In the model, there are $N$ spins. Each spin has two states denoted by \ensuremath{\sigma_{j}=\pm1}, and \ensuremath{J_{jl}} represents the coupling strength between spins $j$ and $l$. The Ising model can be mapped to a variety of combinatorial optimization problems, such as the maximum-cut problem and the traveling-salesman problem. The CIM is constructed by the degenerate optical parametric oscillators (DOPOs) \cite{Jankowski2018PRL,Woo1971IEEE,Okawachi2016OL,Marandi2012OE,Longhi1995OL,Clements2017PRA,Pierangeli2019PRL,Hamerly2019SciAdv,Zhou2021PRA}, where each DOPO unit corresponds to a single spin in the Ising model. Beyond Ising optimization, recent studies have shown that DOPO-based CIMs can generate non-Gaussian resource states, including entangled cat states and coherent cluster states, and can be extended with ancillary modes to remove frustration without altering the target ground-state manifold \cite{Zhou2023PRA,Zhou2025PRL}. For a DOPO, there exists a pump threshold. When the pump amplitude of the system exceeds the pump threshold, the signal field of the DOPO transits from a squeezed state to a coherent state, generating two light fields with phases of 0 and \ensuremath{\pi}, which correspond to the spin states \ensuremath{\sigma_{j}=+1} and \ensuremath{\sigma_{j}=-1} in the Ising model, respectively. Constructing a QB based on such a DOPO system brings two distinct advantages. First of all, the DOPO uses a pump light to drive a second-order nonlinear process within a high-$Q$ resonator to generate signal lights. Due to the high-$Q$ cavity, the intracavity optical field maintains a long coherence time, enabling the CIM-based QB to resist a certain degree of dissipation. Moreover, the DOPO platform is relatively mature. Ever since the optical parametric oscillator was first realized in 1965\cite{Giordmaine1965PRL}, it has been widely applied for squeezed-state generation and CIMs \cite{Wu1986PRL,Yurke1984PRA,Collett1984PRA}.

On the other hand, ergotropy is the maximum amount of work extractable with a unitary transformation. The ergotropy of a QB decays over time due to decoherence and dissipation in the system \cite{Allahverdyan2004EPL,PerarnauLlobet2015PRX}. To clarify the resistance of the different energy components to dissipation, we decompose the ergotropy into its coherent and incoherent parts, which exhibit markedly-different decay rates. By analyzing the evolution of these two components, we can determine the optimal time to switch off the pump field, thus optimizing the energy-storage performance of the QB. In this CIM-based QB, we find that at the pump amplitude 1-1.5 times above the pump threshold, the coherent ergotropy decays roughly half as slowly as the incoherent part and grows significantly earlier. When the coherent component approaches its maximum, the incoherent contribution remains negligible. Remarkably, the average charging power also peaks at nearly the same instant. This coincidence of robustness and charging efficiency provides a clear physical criterion for the optimal switch-off time of the pump field. We further show that the coherent ergotropy increases nonlinearly with pump amplitude and gradually saturates. Finally, coupling to a two-level system (TLS) as the load confirms efficient discharge capability.  
Our work presents the QB architecture on the mature DOPO platform and offers a solid proposal for its experimental realization.

This article is organized as follows. In the next section, we introduce our model. In Sec.~\ref{sec:Charging}, we calculate the ergotropy of the QB based on the master equation. We further investigate the influence of different pump strengths on the total ergotropy as well as on its coherent component. In Sec.~\ref{sec:Discharging}, we study the discharging process to evaluate its performance as an energy source.

\section{Model}
\label{sec:Model}
We consider a charging model of the QB, where the signal mode serves as the battery and the pump mode acts as the charger, as illustrated in Fig.~\ref{fig:1}. Setting $\hbar=1$, the total Hamiltonian is written as~\cite{Wang2013PRA}
\begin{align}\label{eq:Hamiltonian}
H_{\mathrm{tot}}=H_{\mathrm{sys}}+H_{B}+H_{\mathrm{irr}},
\end{align}
where
\begin{align}
H_{\mathrm{sys}}&=H_{0}+H_{\mathrm{int}},\\
H_0 &= \omega_s  a_s^\dagger  a_s + \omega_p  a_p^\dagger  a_p,\\
H_{\mathrm{int}} &= i\frac{\kappa}{2} a_s^{\dagger 2} a_p 
+i\sqrt{\gamma_p} a_p^\dagger F_p e^{-i\omega_p t}+\mathrm{h.c.},\\
H_{B}&=\sum_{j=s,p}\int_{0}^{\infty} d\omega\,
\omega\, b_{j}^{\dagger}(\omega)b_{j}(\omega),\\
H_{\mathrm{irr}}
&=i\sqrt{\gamma_{s}}\, a_{s}^{\dagger} B_{s}
+i\sqrt{\gamma_{p}}\, a_{p}^{\dagger} B_{p}
+\mathrm{h.c.}
\end{align}
Here, $H_0$ describes the free Hamiltonian of the DOPO, and $ a_s^\dagger$ ($ a_p^\dagger$) is the creation operator of the signal (pump) mode with frequency $\omega_s$ ($\omega_p$). The first term in $H_{\mathrm{int}}$ describes the nonlinear interaction between the signal and pump modes, where $\kappa$ denotes the coupling strength responsible for the charging process. The second term represents the coherent driving applied to the pump mode, with $F_p$ being the amplitude of the external classical pump field. Since $F_p$ is the amplitude of the injected classical field, it satisfies $F_p\propto \sqrt{P_{\mathrm{in}}/(\hbar\omega_p)}$~\cite{Yin2009PRA}, where $P_{\mathrm{in}}$ is the input pump power. Treating the injected field as a stable source, we adopt the pump approximation and take $F_p$ to be a constant.
In addition, $b_{j}(\omega)$ and $b_{j}^{\dagger}(\omega)$ are the annihilation and creation operators of the bath mode with frequency $\omega$ in channel $j\in\{s,p\}$. The dissipation Hamiltonian $H_{\mathrm{irr}}$ describes the coupling between the system and the environment, where
\begin{align}
B_{j}=\int_{0}^{\infty} d\omega\, g_{j}(\omega)b_{j}(\omega),
\end{align}
with $g_{j}(\omega)$ being the coupling strength between the system mode and the bath mode at frequency $\omega$. The parameters $\gamma_s$ and $\gamma_p$ characterize the corresponding dissipation rates of the signal and pump modes, respectively.

According to Appendix~\ref{sec:AppendixA}, the master equation of the system can be written as
\begin{align}\label{eq:master}
\frac{d\rho}{dt}
=& -i\bigl[H_{\mathrm{eff}},\rho\bigr]
+\gamma_s\Bigl(a_s\,\rho\,a_s^\dagger
-\tfrac{1}{2}\{a_s^\dagger a_s,\rho\}\Bigr)\nonumber\\
& +\gamma_p\Bigl(a_p\,\rho\,a_p^\dagger
-\tfrac{1}{2}\{a_p^\dagger a_p,\rho\}\Bigr),
\end{align}
Here, the anti-commutator is defined as $\{A,B\}=AB+BA$. 
The first term in Eq.~(\ref{eq:master}) describes the coherent evolution of the QB governed by the effective Hamiltonian $H_{\mathrm{eff}}$. 
The second term characterizes the dissipation of the QB induced by the environment with decay rate $\gamma_s$, while the third term accounts for the dissipative loss of the pump mode (charger) with rate $\gamma_p$. 
The QB is initially prepared in the vacuum state, i.e., $\ket{\psi_s(0)}=\ket{0}$.


In the numerical simulations, we adopt the parameters which are fully-consistent with realistic DOPO. 
In typical experiments \cite{Takata2015PRA,Wang2013PRA,Yamamoto2017npj}, $\gamma_s \ll \gamma_p$ and $\kappa/\gamma_{s}$ ranges from 0.1 to 1. 
The pump amplitude is set to 1--1.5 times the pump threshold $F_{p}^{(th)}$. 
The details are discussed in Appendix~\ref{sec:AppendixB}.
Meanwhile, in the numerical simulation of the master equation, since the photon-number space is infinite, we need to perform a finite-dimensional truncation on it. Under our parameter settings, the truncation dimensions for the signal and pump photon number spaces are taken as $N_{p}=9$ and $N_{s}=32$, respectively. For more details, refer to Appendix~\ref{sec:AppendixC}.

\section{Charging}
\label{sec:Charging}
\begin{figure}
\includegraphics[width=8.5cm]{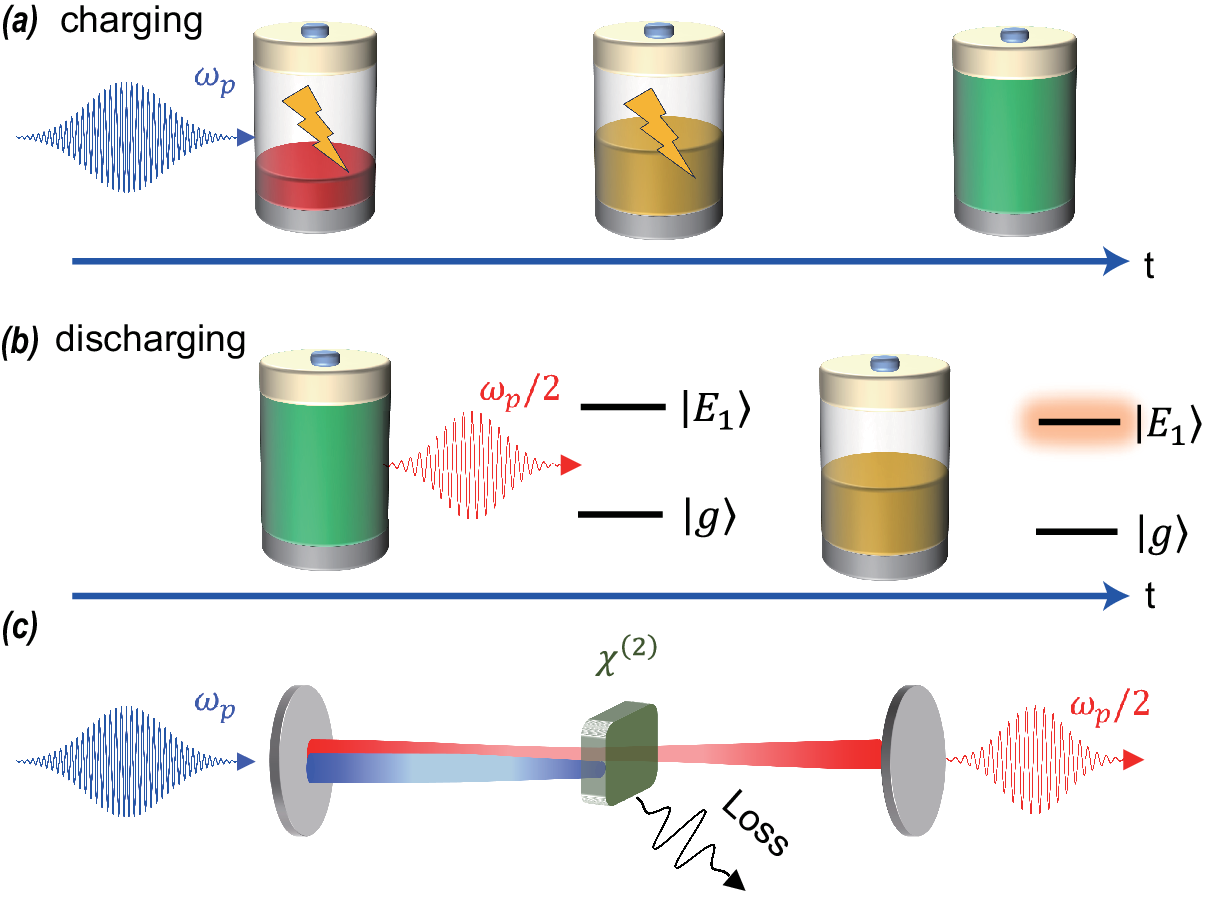}
\caption{Schematic of a CIM-based QB. The entire process is divided into two stages, i.e., (a) charging and (b) discharging. During the charging stage, a pump field with frequency~$\omega_{p}$ is applied to charge the QB. In the discharging stage, after the QB is connected to the load, e.g. a TLS, the QB transfers its stored energy to the TLS, driving it into its excited state. (c) Schematic of a DOPO. A pump field with frequency $\omega_{p}$ is injected into an optical cavity, where it interacts with a second-order nonlinear medium, i.e., $\chi^{(2)}$ crystal, via a three-wave mixing process, generating signal photons at frequency $\omega_{p}/2$. The signal field is in resonance with the cavity, while loss is also present in the system. }\label{fig:1}
\end{figure}

As shown in Fig.~\ref{fig:1}(a), an operation of the QB is divided into two stages, i.e., charging and discharging. During the charging stage, the charger supplies power to the QB. In our model, a pump field at frequency~$\omega_p$ charges the signal mode. In the discharging stage, the QB acts as an energy source, interacting with the atomic system to discharge its stored excitation.

When evaluating the performance of a QB, we need to focus on the energy stored in the QB, which is given by $\textrm{Tr}\left[\rho_{s}{H}_{S}\right]$. However, according to the second law of thermodynamics, not all of this energy can be extracted and converted into useful work. Here, we utilize ergotropy to quantify the amount of extractable work stored in the QB, which serves as one of the key indicators of QB's performance \cite{Allahverdyan2004EPL,PerarnauLlobet2015PRX}. It is defined as
\begin{equation}\label{eq:5}
W(t)
= \mathrm{Tr}\bigl[\rho_{s}\,H_{S}\bigr]
- \mathrm{Tr}\bigl[\widetilde{\rho}_{s}\,H_{S}\bigr],
\end{equation}
where $H_{S}=\omega_s  a_s^\dagger  a_s=\sum_{n=0}^{n=\infty} n\omega_s\ket{n}_{ss}\!\!\bra{n}$ is the Hamiltonian of the QB, $\widetilde{\rho}_{s}=\sum_n r_n \ket{n}_{ss}\!\!\bra{n}$ is the passive
state of the QB, obtained by rearranging the eigenvalues of $\rho_{s}$ in descending order. The passive state is a quantum state at which no work can be extracted via any unitary transformation. In Eq.~(\ref{eq:5}), the first term represents the total energy stored in the QB, while the second term corresponds to the energy of the passive state, i.e., the portion of energy that can not be extracted as useful work.

\begin{figure}
\includegraphics[width=8.5cm]{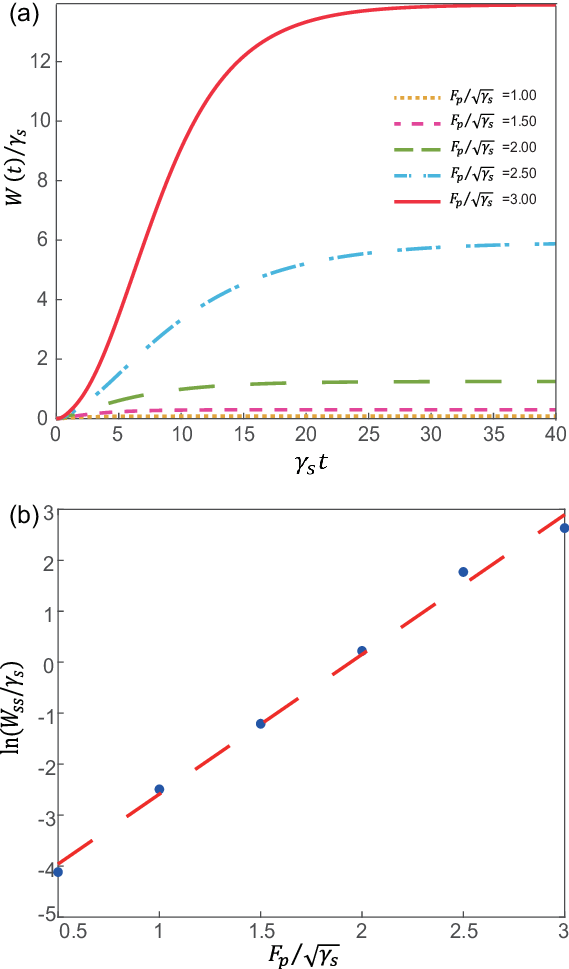}
\caption{(a) Time evolution of the QB's ergotropy under different $F_{p}/\sqrt{\gamma_{s}}$s, where the yellow dotted line, the pink short-dashed line, the green long-dashed line, the blue dash-dotted line, and the red solid line correspond to $F_{p}/\sqrt{\gamma_{s}} = 1.00,\, 1.50,\, 2.00,\, 2.50,$ and $3.00$, respectively. (b) The steady-state ergotropy $W_{ss}$ of the QB for different $F_{p}/\sqrt{\gamma_{s}}$s is fitted linearly as $\ln (W_{ss}/\gamma_{s}) = -5.330 + 2.742\times F_{p}/\sqrt{\gamma_{s}}$, yielding a correlation coefficient of $|r| = 0.9974$. We use the following parameters, i.e., $\Delta = 0$, $\kappa/\gamma_s = 0.5$, and $\gamma_p/\gamma_s = 16$.
}\label{fig:2}
\end{figure}

In Fig.~\ref{fig:2}(a), the ergotropy gradually approaches a steady state. Moreover, it can be observed that the steady-state value of ergotropy $W_{ss}/\gamma_{s}$ increases with $F_{p}/\sqrt{\gamma_{s}}$. Figure~\ref{fig:2}(b) shows that $W_{ss}/\gamma_{s}$ grows exponentially with $F_{p}/\sqrt{\gamma_{s}}$ approximately. 
In the DOPO system, below the pump threshold, although the steady-state mean field of the signal mode satisfies $\alpha_{s}(\infty)=0$, quantum squeezing fluctuations still exist \cite{Wang2013PRA}. As $F_{p}/\sqrt{\gamma_{s}}$ increases, these squeezing fluctuations gradually become stronger, reaching their maximum near the threshold, and then diminish beyond it. Above the pump threshold, the signal mode evolves into a quasi-classical coherent state. Therefore, $F_{p}/\sqrt{\gamma_{s}}$ does not only significantly affect the magnitude of the QB's ergotropy, but also strongly influences the proportion of coherent and squeezed light stored within the QB.

We can further divide the ergotropy into the incoherent part $W^{i}(t)$ and the coherent part $W^{c}(t)$ \cite{Song2025PRL}, i.e.,
\begin{align}
W(t)=W^{i}(t)+W^{c}(t).
\end{align}
Since the off-diagonal elements of the density matrix represent quantum coherence, 
$W^{i}(t)$ and $W^{c}(t)$ can be explicitly written as \cite{Song2025PRL}
\begin{align}
W^{i}(t)	&=\mathrm{Tr}[{H}_{s}\rho_{s}(t)]-\mathrm{Tr}[{H}_{s}\tilde{\varrho}_{s}(t)],\\
W^{c}(t)	&=\mathrm{Tr}[{H}_{s}\tilde{\varrho}_{s}(t)]-\mathrm{Tr}[{H}_{s}\tilde{\rho}_{s}(t)].
\end{align}
Here, $\varrho_{s}(t)$ denotes the dephased state of $\rho_{s}(t)$ in the eigen-energy basis $|n\rangle_{s}$, obtained by removing all off-diagonal elements, i.e., $\varrho_{s}(t) = \sum_{ns}\!\langle n|\rho_{s}(t)|n\rangle_{s}\ket{n}_{ss}\!\!\bra{n}$,
while $\tilde{\varrho}_{s}(t)$ is the passive state constructed from $\varrho_{s}(t)$ by rearranging its eigenvalues $\lambda_{n}$s in descending order and $\ket{n}_{s}$s in ascending order, i.e., $\tilde{\varrho}_{s}(t)=\sum_{n}\lambda_{n}\ket{n}_{ss}\!\!\bra{n}$.

\begin{figure}
\includegraphics[width=8.5cm]{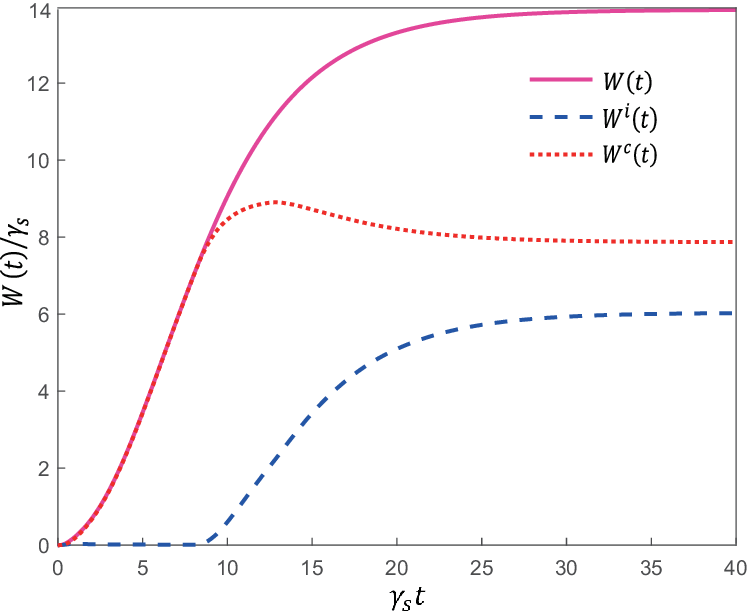}
\caption{The time evolution of the ergotropy $W(t)$, and its coherent part $W^{c}(t)$, and its incoherent part $W^{i}(t)$. The pink solid, red dotted, and blue dashed lines correspond to $W(t)$, $W^{c}(t)$, and $W^{i}(t)$, respectively.  We use the following parameters, i.e., $\Delta = 0$, $\kappa/\gamma_s = 0.5$, $\gamma_p/\gamma_s = 16$, and $F_{p}/\sqrt{\gamma_{s}} = 3.0$.}\label{fig:3}
\end{figure}

As shown in Fig.~\ref{fig:3}, during the initial stage of QB charging, i.e., when $\gamma_{s}t$ ranges from 0 to 8, the increase in ergotropy entirely originates from the coherent part. Around $\gamma_{s}t=10$, the coherent ergotropy reaches its maximum. Afterwards followed by a slight decrease, it finally approaches a steady state. In contrast, the incoherent ergotropy starts to grow only after $\gamma_{s}t>8$, and eventually reaches its own steady state. 

\begin{figure}
\includegraphics[width=8.5cm]{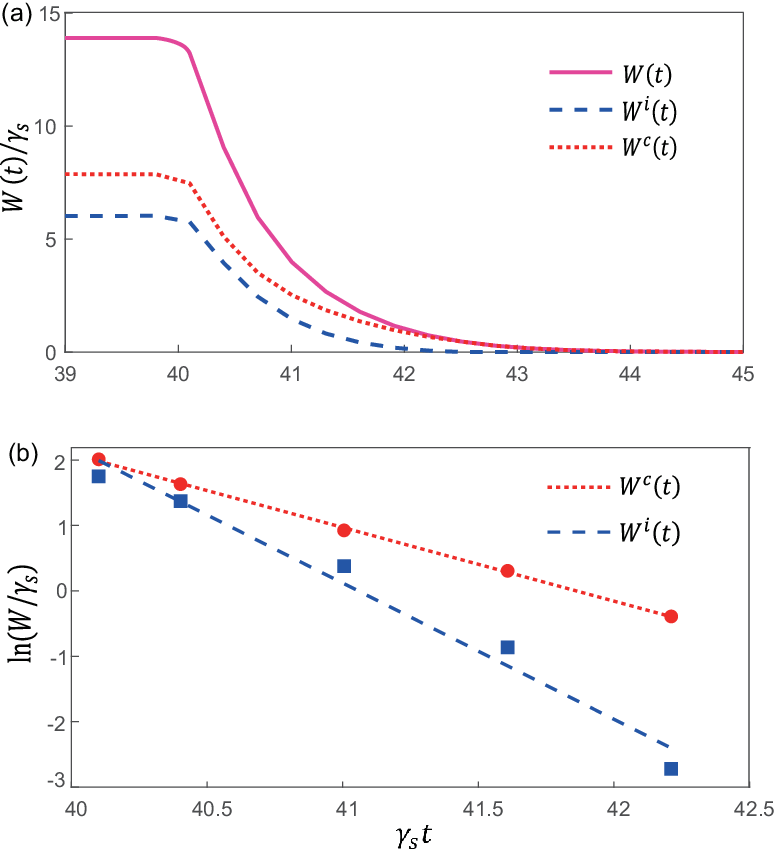}
\caption{(a) Time evolution of the ergotropy $W(t)$, and its coherent part $W^{c}(t)$, 
  and its incoherent part $W^{i}(t)$ when $F_{p}/\sqrt{\gamma_{s}}=0$ since $\gamma_{s}t=40$. 
  The pink solid, red dotted, and blue dashed lines correspond to $W(t)$, $W^{c}(t)$, and $W^{i}(t)$, respectively. (b) $W^{c}(t)$ ($W^{i}(t)$) by the red circles (the blue squares) are linearly fitted by 
  $\ln (W^{c}(t)/\gamma_{s})=-1.127\times\gamma_{s}t+47.17$ ($\ln (W^{i}(t)/\gamma_{s})=-2.082\times\gamma_{s}t+85.48$)
  with the correlation coefficient $|r^{c}|=0.9996$ ($|r^{i}|=0.9881$).
   We use the following parameters, i.e., $\Delta = 0$, $\kappa/\gamma_s = 0.5$, $\gamma_p/\gamma_s = 16$, and $F_{p}/\sqrt{\gamma_{s}} = 3.0$ before $\gamma_{s}t=40$.
  }\label{fig:4}
\end{figure}

\begin{figure}
\includegraphics[width=8.5cm]{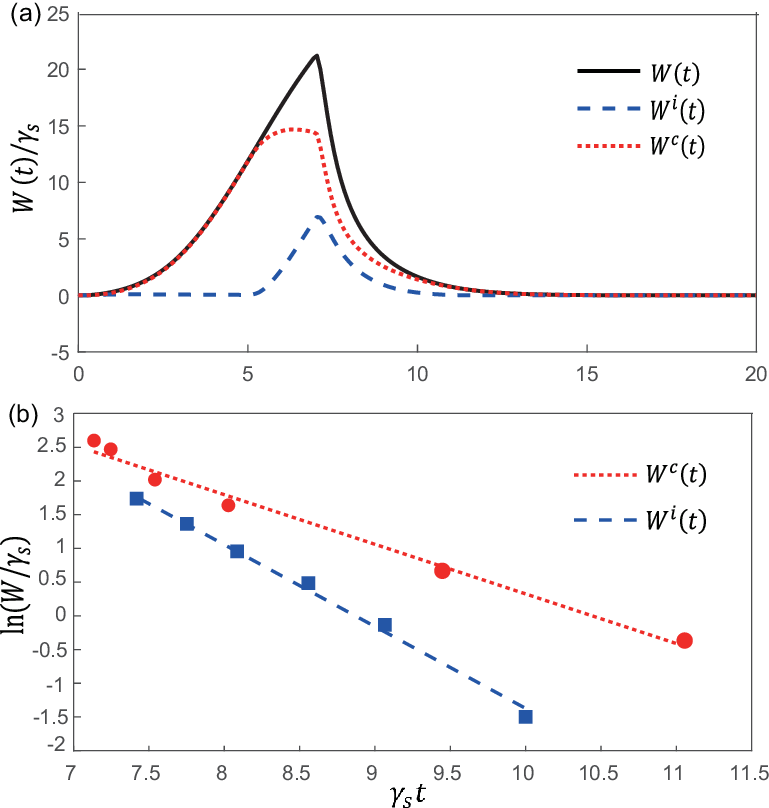}
\caption{(a) Time evolution of the ergotropy $W(t)$, and its coherent part $W^{c}(t)$, 
  and its incoherent part $W^{i}(t)$ when $F_{p}/\sqrt{\gamma_{s}}=0$ since $\gamma_{s}t=7$. 
  The black solid, red dotted, and blue dashed lines correspond to $W(t)$, $W^{c}(t)$, and $W^{i}(t)$, respectively. (b) $W^{c}(t)$ ($W^{i}(t)$) by the red circles (the blue squares) are linearly fitted by 
  $\ln (W^{c}(t)/\gamma_{s})=-0.7361\times\gamma_{s}t+7.6856$ ($\ln (W^{i}(t)/\gamma_{s})=-1.2146\times\gamma_{s}t+10.7728$)
  with the correlation coefficient $|r^{c}|=0.9881$ ($|r^{i}|=0.9972$). The parameters are $N_{s}=60$, $N_{p}=9$, $\Delta=0$, $\gamma_{p}/\gamma_{s}=32$, $\kappa/\gamma_s=1.00$, and $F_{p}/\sqrt{\gamma_{s}}=4.24
  $.}\label{fig:5}
\end{figure}

A key challenge for QBs is self-discharging. Once the drive $F_{p}$ is switched off, the ubiquitous decoherence inevitably forces the stored ergotropy to decay spontaneously.  To investigate the resistance of the coherent and incoherent ergotropy against decoherence, we set $F_{p}=0$ at $\gamma_{s}t=40$ to simulate the evolution of the ergotropy with respect to $t$ after turning off the pumping field. As shown in Fig.~\ref{fig:4}, after the pump field is switched-off, it is observed that both $W^{c}/\gamma_{s}$ and $W^{i}/\gamma_{s}$ decay gradually to zero, while $W^{c}/\gamma_{s}$ exhibits a slower decay rate. When performing a linear fit of $\ln W/\gamma_{s}$ versus $\gamma_{s}t$, we can obtain $\ln (W^{c}(t)/\gamma_{s})=-1.127\times\gamma_{s}t+47.17$ and 
$\ln (W^{i}(t)/\gamma_{s})=-2.082\times\gamma_{s}t+85.48$. The decay rate of the incoherent ergotropy is approximately twice that of the coherent ergotropy. This indicates that, in our QB, the coherent ergotropy exhibits a more robust character and can effectively resist decoherence. The dissipation of the system can be suppressed by employing a high-$Q$ cavity. As shown in Fig.~\ref{fig:5}, our results demonstrate that when optical cavities with a larger $Q$-factor are used, the decay rates of both the coherent and incoherent parts of ergotropy are significantly reduced compared with the original configuration. Meanwhile, the decay rate of the incoherent part remains faster than that of the coherent part, which further confirms that the coherent component exhibits a stronger decay-resistant characteristic. The incoherent part of ergotropy is governed by population inversion, whereas the coherent part originates from the off-diagonal coherence of the density matrix. This leads to different behaviors for the two contributions. In Appendix~\ref{sec:AppendixD}, taking a two-level QB as an example, we show the decay of the incoherent and coherent parts of ergotropy explicitly.

\begin{figure}
\includegraphics[bb=25 0 410 600, width=8.5cm]{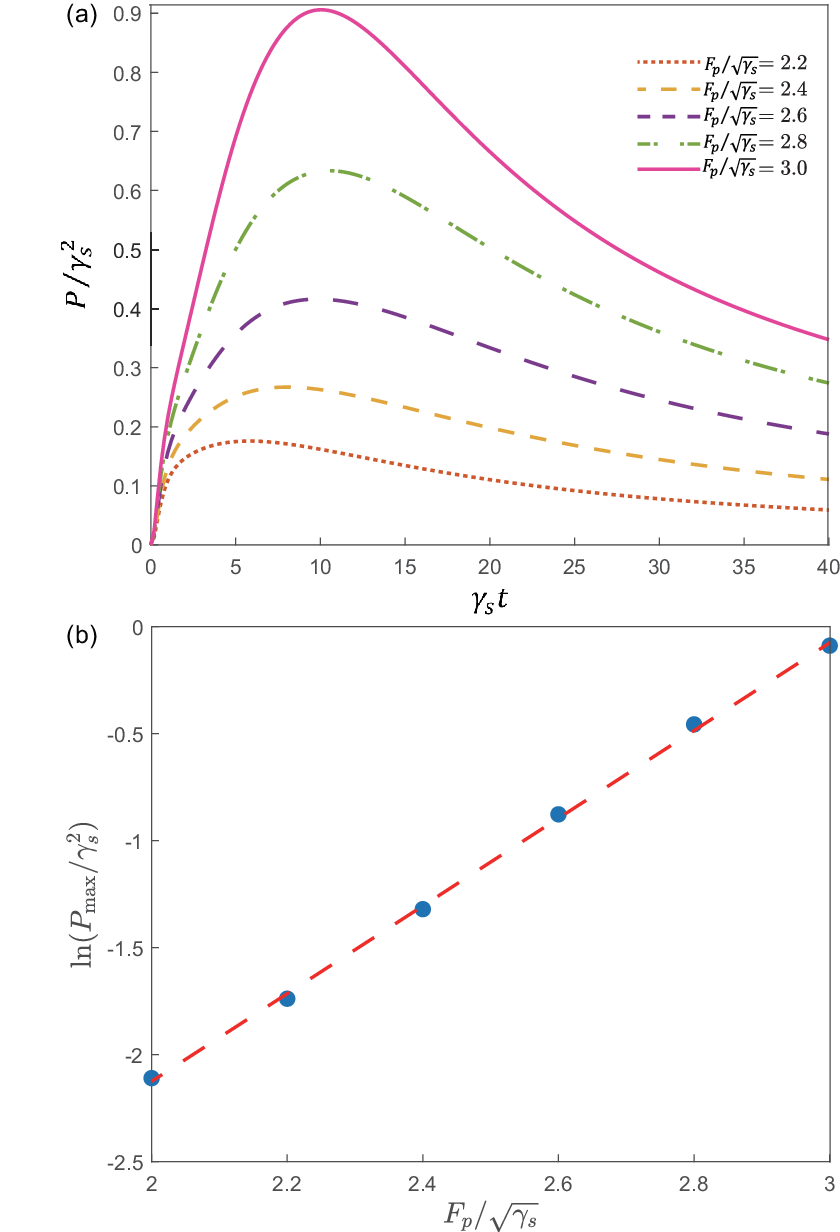}
\caption{(a) The average charging power $P$ of the QB vs different $F_{p}/\sqrt{\gamma_{s}}$, where the red dotted, the yellow short-dashed, the purple long-dashed, the green dash-dotted, and the pink solid lines correspond to $F_{p}/\sqrt{\gamma_{s}} = 2.2,\, 2.4,\, 2.6,\, 2.8,\, 3.0$, respectively. (b) The variation of $\ln(P_{\max}/\gamma_{s}^{2})$ with $F_{p}/\sqrt{\gamma_{s}}$ is fitted as
$\ln(P_{\max}/\gamma_{s}^{2}) = 2.049 \times F_{p}/\sqrt{\gamma_{s}} - 6.222$.
The correlation coefficient is $|r|=0.9995$. We use the following parameters, i.e., $\Delta = 0$, $\kappa/\gamma_s = 0.5$, $\gamma_p/\gamma_s = 16$, and $F_{p}/\sqrt{\gamma_{s}}=3.00$.
}\label{fig:6}
\end{figure}

Then, we discuss the average charging power of the QB, which is defined as $P=W(t)/t$ \cite{Ferraro2018PRL}. It reflects the rate at which the QB is being charged. A higher average power indicates a faster charging rate, while a lower power suggests a slower charging rate. As shown in Fig.~\ref{fig:6}(a), we observe that as $F_{p}/\sqrt{\gamma_{s}}$ increases, the average power also raises, indicating that a larger $F_{p}/\sqrt{\gamma_{s}}$ accelerates the charging process of the QB. However, after reaching its maximum, the average power begins to decrease over time, which suggests that the rate of increase in ergotropy slows down as the QB approaches its maximum charging capacity. As shown in Fig.~\ref{fig:6}(b), with the parameters employed in this work, $P_{\max}/\gamma_{s}^{2}$ increases exponentially with $F_{p}/\sqrt{\gamma_{s}}$. This behavior indicates that enhancing the driving-field strength can significantly boost the maximum charging power of the QB.

In the QBs made of $N$ TLSs, the improvement in performance is usually described by how it scales with $N$, rather than with an external driving parameter. Depending on the model and regime, the charging power can show different scaling behaviors with $N^2$. In contrast, in our CIM-based QB, both the ergotropy and the charging power increase exponentially with the pump amplitude $F_p$ above the threshold \cite{Rouse2022SA}. This shows a different way to enhance battery performance, coming from the nonlinear pump response of the CIM platform rather than the usual size scaling.

Combining Figs.~\ref{fig:4} and \ref{fig:6}, we observe that both the maximum coherent ergotropy and the average charging power peak around $\gamma_s t \approx 10$. When the pump is switched off at this moment, compared to charging the QB to the steady state at $\gamma_s t = 30$, although the maximum stored energy is somewhat reduced, the amount of the more decay-resistant coherent ergotropy is instead significantly higher. Moreover, the charging time is shortened by a factor of 3, which means that the number of charge-discharge cycles for the QB can be significantly increased within the same duration.

\begin{figure}
\includegraphics[width=8.5cm]{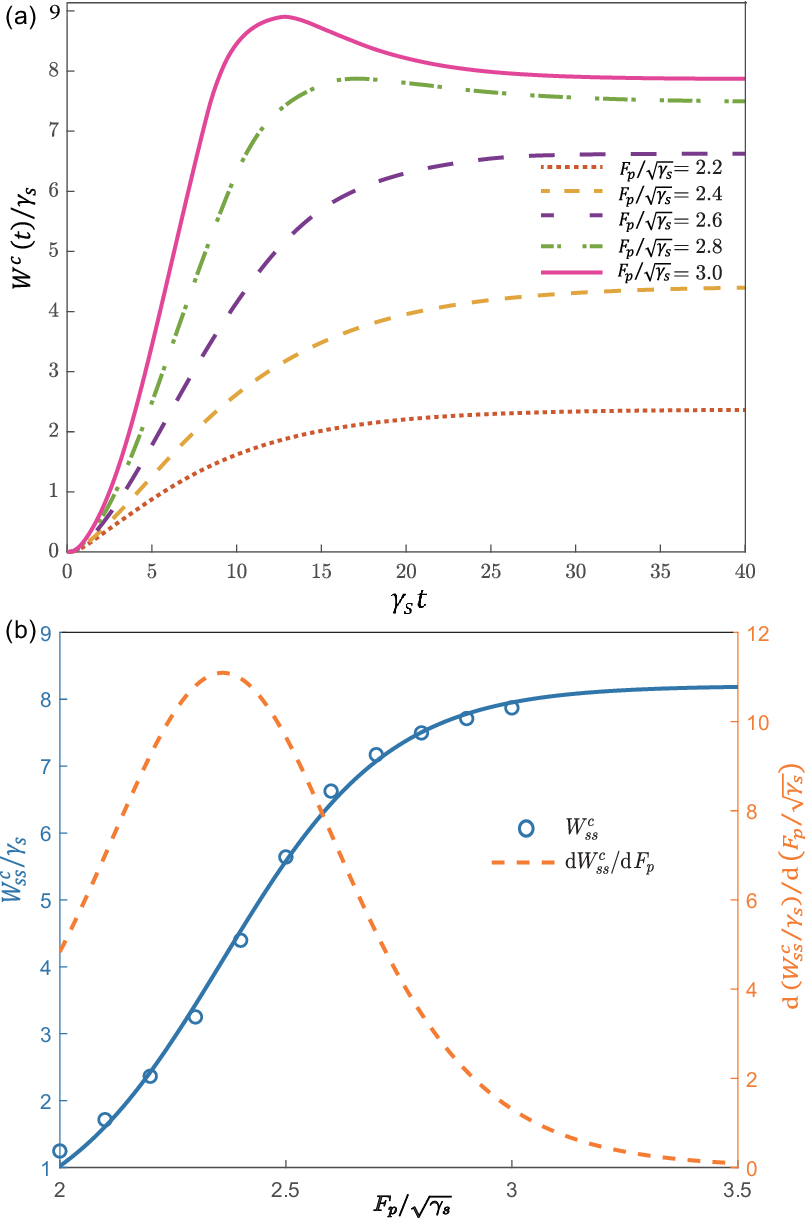}
\caption{(a) Time evolution of the QB's coherent part $W^{c}(t)$ under different $F_{p}/\sqrt{\gamma_{s}}$s, where the red dotted, the yellow short-dashed, the purple long-dashed, the green dash-dotted, and the pink solid lines correspond to $F_{p}/\sqrt{\gamma_{s}} = 2.2,\, 2.4,\, 2.6,\, 2.8,\, 3.0$, respectively. (b) The steady-state values of $W^c$ are evaluated for $F_{p}/\sqrt{\gamma_{s}} \in [2.0, 3.5]$ with an increment of 0.1, shown as the blue solid line. The derivative $\mathrm{d}W^c/\mathrm{d}F_p$ is computed numerically and represented by the orange dashed line. We use the following parameters, i.e., $\Delta = 0$, $\kappa/\gamma_s = 0.5$, and $\gamma_p/\gamma_s = 16$. }\label{fig:7}
\end{figure}

As shown in Fig.~\ref{fig:7}(a), the quantum dynamics of the coherent part $W^{c}(t)/\gamma_{s}$ is significantly tuned by the pump strength $F_{p}/\sqrt{\gamma_{s}}$. When $F_{p}/\sqrt{\gamma_{s}}$ is small, $W^{c}(t)/\gamma_{s}$ increases monotonically with respect to the time and finally reaches its steady state. However, as $F_{p}/\sqrt{\gamma_{s}}$ is larger than 2.7, which is not shown here, $W^{c}(t)/\gamma_{s}$ will be increased to its maximum, followed by a small decrease to its steady state.
These observations indicates that the driving field sensitively influences the ordered energy stored in the QB. Then, we further investigate the steady state of $W^{c}(t)/\gamma_{s}$ in Fig.~\ref{fig:7}(b). By numerically fitting, we can obtain the following expression 
\begin{equation}
W_{ss}^{c}/\gamma_{s}=\frac{8.2016}{1+\exp[-5.4097\times\left(\frac{F_{p}}{\sqrt{\gamma_{s}}}-2.3605\right)]}.
\end{equation} 
Therein, although the steady-state value of $W^{c}(t)/\gamma_{s}$ exhibits a monotonic dependence on $F_{p}/\sqrt{\gamma_{s}}$, which suggests that $W^{c}_{ss}/\gamma_{s}$ approaches 8.0 in the strong-driving-field limit, its derivative with respect to $F_{p}/\sqrt{\gamma_{s}}$ shows its maximum around $F_{p}/\sqrt{\gamma_{s}} =2.4$. All these discoveries imply that once the pump exceeds the threshold by a certain margin, the system enters a nonlinear-limited regime in which additional input energy can no longer be efficiently converted into coherent ergotropy.
In other words, in a CIM-based QB, there exists an optimal operation point for the pump field. By appropriately selecting $F_{p}/\sqrt{\gamma_{s}}$, one can obtain a relatively-large $W^{c}_{ss}/\gamma_{s}$ while simultaneously saving input energy, thereby improving the overall energy-utilization efficiency.

Beyond the CIM-based QB studied here, it would be interesting to explore in a CIM whether the remaining intracavity energy can be reused as extractable energy after an calculation has been performed. A detailed study of this question, including network effects, output readout, and energy transfer to an external load, may be explored for the future work.

\section{Discharging}
\label{sec:Discharging}

To investigate the discharge behavior, we couple the QB to a TLS as a load \cite{Catalano2024PRX}. The TLS is a fundamental quantum system and is widely employed in quantum optics and quantum information as a prototype for qubits. Using a TLS does not only simplify theoretical analysis but also links our model to practical quantum devices.
The system Hamiltonian is written as
\begin{equation}
{H}' = \omega_{s}\,{a}_{s}^{\dagger}{a}_{s}
   + \frac{\omega_{a}}{2}\,{\sigma}_{z}
   + g\bigl({a}_{s}\,{\sigma}_{+}+{a}^{\dagger}_{s}\,{\sigma}_{-}\bigr),
\end{equation}
where $\sigma_{z}$ is the Pauli operator of the TLS, $\sigma_{+}$ ($\sigma_{-}$) represents the raising (lowering) operator of the TLS with the level spacing $\omega_a$, $g$ represents the coupling strength between QB and TLS. 
Here, $\sigma_{+}$ excites the atom from its ground state $\vert g\rangle$ to the excited state $\vert e\rangle$, 
when a photon from the QB is absorbed.

On account of the dissipation of both the QB and the TLS, 
the time evolution of the total density matrix $\rho'$ is governed by 
the Lindblad-form quantum master equation \cite{breuer2002}
\begin{align}
\frac{d{\rho}'}{dt} = & -i\bigl[{H}',\,{\rho}'\bigr]  + \gamma_{s}\Bigl({a}_{s}\,{\rho}'\,{a}^{\dagger}_{s}
      - \tfrac{1}{2}\{{a}_{s}^{\dagger}{a}_{s},\,{\rho}'\}\Bigr) \nonumber \\
& + \gamma_{a}\Bigl({\sigma}_{-}\,{\rho}'\,{\sigma}_{+}
      - \tfrac{1}{2}\{{\sigma}_{+}{\sigma}_{-},\,{\rho}'\}\Bigr),
\label{eq:discharge_master}
\end{align}
where $\gamma_{s}$ and $\gamma_{a}$ denote the relaxation rates of the 
QB and the TLS during discharging, respectively. 
The first term describes unitary evolution under $H'$, 
and the second term accounts for the photon loss of the QB, 
and the third term represents the atomic dissipation.

When the driving is turned off, e.g. $\gamma_{s}t_0 = 10$, 
the TLS is initialized in its ground state, and the density matrix of the QB and the TLS reads
\begin{equation}
\rho'(t_0)
 = \rho_{s}(t_0)\;\otimes\;\vert g\rangle\langle g\vert
\end{equation}
where $\rho_{s}(t_0)$ denotes the reduced density matrix of the QB at the instant 
of disconnection with the charger. By solving Eq.~(\ref{eq:discharge_master}) 
and tracing over the atomic degrees of freedom, 
the reduced density matrix of the QB and the TLS are obtained as
\begin{align}
\rho_{s}(t) = \mathrm{Tr}_{\mathrm{a}}\!\left[\rho'(t)\right],\\
\rho_{a}(t) = \mathrm{Tr}_{\mathrm{s}}\!\left[\rho'(t)\right].
\end{align}

\begin{figure}
\includegraphics[width=8.5cm]{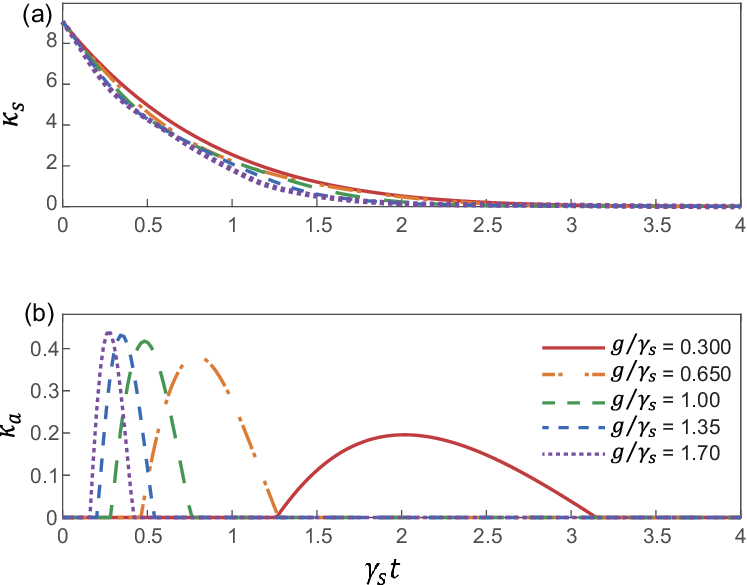}
\caption{The time evolution of (a) $\kappa_{s}$ and (b) $\kappa_{a}$ during the discharging process, where the red solid, the orange dash-dotted, the green long-dashed, the blue short-dashed, and the purple dotted lines correspond to $g/\gamma_{s} = 0.300,\, 0.650,\, 1.00,\, 1.35,\, 1.70$, respectively. The other parameters are $\omega_{s}/\gamma_{s}=100$, $\omega_{a}/\gamma_{s}=100$, and $\gamma_{a}/\gamma_{s}=0.1$. The discharging process starts from the moment when the pump field is switched off at $\gamma_{s} t_{0}=10$.}\label{fig:8}
\end{figure}

By Eq.~(\ref{eq:5}), we can calculate the time evolution 
of the ergotropy $W_{s}$ ($W_{a}$) of the QB (TLS) during the 
discharging process. 
Here, we focus on the normalized ergotropy of the QB and the TLS \cite{Catalano2024PRX}, which is defined as
\begin{equation}
\kappa_{i}=W_{i}/\omega_{i}\ (i=s,a).
\end{equation}

As shown in Fig.~\ref{fig:8}, when the coupling strength $g$ between the QB and the TLS is gradually increased, the normalized ergotropy, which the load can ultimately obtain, exhibits pronounced non-monotonic behavior.
In the weak-coupling regime, e.g., $g/\gamma_{s} = 0.300$, the peak value of $\kappa_a$ is relatively low and rises very slowly. This is because the smaller the coupling strength $g/\gamma_{s}$ is, the longer the  period of the Rabi oscillations between the QB and the TLS becomes. As a consequent, the time required to complete efficient energy transfer increases accordingly. Meanwhile, the dissipation in the system continuously causes energy loss, leading to a decay over time in the ergotropy stored in the QB. Since the energy transfer takes a certain amount of time to complete, before the TLS absorbs sufficient energy from the QB, the ergotropy of the QB has already been significantly reduced due to the persistent dissipation, ultimately resulting in a lower peak value of $\kappa_{a}$ and a slow rising process.
As $g/\gamma_{s}$ increases, the energy transfer accelerates dramatically, and thus increases the peak value of $\kappa_a$ and reduces the time required to reach it markedly. Once $g/\gamma_{s} \gtrsim 1.00$, the maximum $\kappa_a$ saturates at approximately $0.43$ and stops to grow further with increasing $g$. This saturation indicates that the transfer efficiency is fundamentally limited by the finite ergotropy $W_s(t_0)$ stored in the QB at the instant switching off the pump, rather than by the coupling strength itself.


A particularly-important observation is the existence of an optimal discharging time. For a given $g$, $\kappa_a(t)$ exhibits a Rabi-like oscillation. $\kappa_a(t)$ reaches maximum at the first peak, with later peaks decaying rapidly due to the accumulative dissipation during each energy exchange. Therefore, the highest energy-transfer efficiency is achieved by disconnecting the load with the QB precisely at the first maximum of $\kappa_a(t)$.

Combining this insight with the charging stage analysis, particularly Figs.~\ref{fig:4} and~\ref{fig:6}, $W_c(t)$ reaches its maximum near $\gamma_{s} t \approx 10$ during the charging stage, and this nearly coincides with the maximum of the average charging power. Therefore, $\gamma_{s} t \approx 10$ defines the optimal time to switch off the pump. Connecting a TLS with a strong coupling exactly at this moment therefore offers the prospect of approaching the upper limit of discharge efficiency.

\begin{figure}[h]
\includegraphics[width=8.5cm]{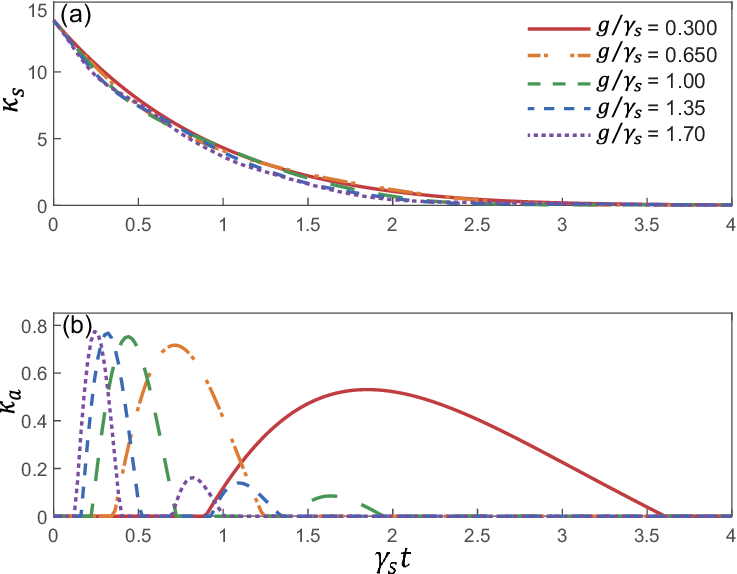}
\caption{The time evolution of (a) $\kappa_{s}$ and (b) $\kappa_{a}$ during the discharging process, where the red solid, the orange dash-dotted, the green long-dashed, the blue short-dashed, and the purple dotted lines correspond to $g/\gamma_{s} = 0.300,\, 0.650,\, 1.00,\, 1.35,\, 1.70$, respectively. The other parameters are $\omega_{s}/\gamma_{s}=100$, $\omega_{a}/\gamma_{s}=100$, and $\gamma_{a}/\gamma_{s}=0.1$. The discharging process starts from the moment when the pump field is switched off at $\gamma_{s} t_{0}=30$.}\label{fig:9}
\end{figure}

We further observe that $\kappa_a$ is closely related to the ergotropy stored in the QB. As the charging time is extended to $\gamma_{s} t_0 = 30$, $\kappa_a$ increases to $0.8$, as shown in Fig.~\ref{fig:9}. This value is almost twice as that reported in Ref.~\cite{Catalano2024PRX}, indicating that the corresponding discharging performance is significantly improved.

Our scheme can be simply implemented on a CIM platform, with a signal mode as the QB and the pump field for charging. The battery can be charged by increasing the pump amplitude above the threshold, and the charging process can in principle be stopped at any time. In particular, the peak of the coherent ergotropy provides a useful indication of the optimal charging time. For the discharging stage, the stored energy may be transferred to a TLS acting as a load through a controllable light-matter coupling, and the discharge performance can be characterized by monitoring the population change and the corresponding normalized ergotropy of the load. The main obstacle in such an implementation would include the cavity loss, pump fluctuations, imperfect switching of the pump field, decoherence of the load, and inefficiency in the energy-transfer coupling. Nevertheless, since the dissipation is already included in our model, the present results provide a useful starting point for assessing the experimental feasibility of this scheme.

\section{Conclusion}
\label{sec:Conclusion}

In this work, we propose a QB based on the DOPO. Inside the cavity, the pump field acts as the charger, while the signal field serves as the QB to store energy. The two fields are coupled via a nonlinear crystal, enabling energy transfer from the pump to the signal field. By employing the master equation, we calculate the ergotropy under different parameters and investigate how the driving strength $F_{p}$ influences the steady-state ergotropy. The results show that the steady-state ergotropy increases exponentially with $F_{p}$. 

We further analyze the coherent and incoherent contributions to the ergotropy and find that the coherent component exhibits stronger resistance to the dissipation. Interestingly, the coherent ergotropy does not increase indefinitely with $F_{p}$. Instead, it saturates at an upper limit, revealing an optimal driving strength for efficient energy utilization. Furthermore, based on the time evolution of both the coherent ergotropy and the average charging power, we identify the optimal pump switch-off instant for the QB. Finally, by coupling a TLS to the QB, we examine its discharging behavior and confirm that the CIM-based QB can effectively release its stored energy to a quantum load. By precisely controlling the times for switching off the pump field and connecting and disconnecting the load with the QB, a highly-efficient and fully-controllable charging and discharging cycle of a QB can be realized.

Overall, our proposed CIM-based QB does not only provide an experimentally-feasible platform but also enriches the family of quantum energy-storage devices. This work lays the foundation for future implementations of high-performance QB with enhanced robustness against decoherence. 

\section{Acknowledgments}

This work is supported by the National Natural Science Foundation of China under Grant No.~62461160263, and Quantum Science and Technology-National Science and Technology Major Project (2023ZD0300200), and Guangdong Provincial Quantum Science Strategic Initiative under Grant No.~GDZX2505004.

\appendix

\section{Derivation of the Finite-temperature Master Equation}
\label{sec:AppendixA}

In this appendix, we derive the master equation of the system \cite{Gardiner1985PRA}.
The total Hamiltonian is
\begin{equation}
H_{\mathrm{tot}}=H_{\mathrm{sys}}+H_{\mathrm{B}}+H_{\mathrm{irr}},
\end{equation}
where $H_{\mathrm{tot}}$ is the total Hamiltonian, including the system and the bath.
Here,
\begin{equation}
H_{\mathrm{sys}}=H_{0}+H_{\mathrm{int}},
\end{equation}
where $H_{\mathrm{sys}}$ describes the DOPO system itself, $H_0$ is the free Hamiltonian, and $H_{\mathrm{int}}$ is the interaction Hamiltonian.
With
\begin{equation}
H_{0}=\omega_{s}a_{s}^{\dagger}a_{s}+\omega_{p}a_{p}^{\dagger}a_{p},
\end{equation}
where $a_s^\dagger$ ($a_s$) and $a_p^\dagger$ ($a_p$) are the creation (annihilation) operators of the signal and pump modes with frequencies $\omega_s$ and $\omega_p$, respectively.
And
\begin{equation}
H_{\mathrm{int}}
=i\frac{\kappa}{2}a_{s}^{\dagger 2}a_{p}
+i\sqrt{\gamma_{p}}\,a_{p}^{\dagger}F_{p}e^{-i\omega_{p}t}
+\mathrm{h.c.},
\end{equation}
where the first term describes the nonlinear coupling between the signal and pump modes, with $\kappa$ being the corresponding coupling strength. And the second term represents the coherent driving on the pump mode, where $F_p$ is the amplitude of the external classical pump field and $\gamma_p$ is the dissipation rate of the pump mode.
The bath Hamiltonian is
\begin{equation}
H_{\mathrm{B}}=\sum_{j=s,p}\int_{0}^{\infty} d\omega\,
\omega\, b_{j}^{\dagger}(\omega)b_{j}(\omega).
\end{equation}
Here, $b_j^\dagger(\omega)$ ($b_j(\omega)$) is the creation (annihilation) operator of the bath mode with frequency $\omega$ in channel $j\in\{s,p\}$.
The system-bath interaction is
\begin{equation}
H_{\mathrm{irr}}
=i\sqrt{\gamma_{s}}\,a_{s}^{\dagger}B_{s}
+i\sqrt{\gamma_{p}}\,a_{p}^{\dagger}B_{p}
+\mathrm{h.c.},
\end{equation}
where $\gamma_s$ and $\gamma_p$ characterize the dissipation rates of the signal and pump modes, respectively.
With
\begin{equation}
B_{j}=\int_{0}^{\infty} d\omega\, g_{j}(\omega)b_{j}(\omega).
\end{equation}
Here, $B_j$ is the collective bath operator for channel $j\in\{s,p\}$, and $g_j(\omega)$ denotes the coupling strength between the system mode and the bath mode at frequency $\omega$.
Furthermore, we assume the resonance condition
\begin{equation}
2\omega_{s}=\omega_{p}.
\end{equation}

To derive the reduced dynamics, we move to the interaction picture with respect to the free Hamiltonian $H_0+H_B$.
The system operators evolve as
\begin{align}
a_{j}(t)&=a_{j}e^{-i\omega_{j}t},\\
a_{j}^{\dagger}(t)&=a_{j}^{\dagger}e^{i\omega_{j}t},
\end{align}
while the bath operators evolve as
\begin{equation}
b_{j}(\omega,t)=b_{j}(\omega)e^{-i\omega t}.
\end{equation}
Accordingly,
\begin{equation}
B_{j}(t)=\int_{0}^{\infty} d\omega\, g_{j}(\omega)b_{j}(\omega)e^{-i\omega t}.
\end{equation}

Under the resonance condition, the interaction Hamiltonian in the interaction picture becomes time independent as
\begin{equation}
H_{\mathrm{eff}}
=
i\frac{\kappa}{2}\left(a_{s}^{\dagger 2}a_{p}-a_{s}^{2}a_{p}^{\dagger}\right)
+i\sqrt{\gamma_{p}}\left(F_{p}a_{p}^{\dagger}-F_{p}^{\ast}a_{p}\right).
\end{equation}
For the sake of simplicity, we assume $F_p$ to be real.
The system-bath coupling in the interaction picture is
\begin{equation}
H_{\mathrm{irr}}^{\mathrm{I}}(t)
=
i\sum_{j=s,p}\sqrt{\gamma_{j}}
a_{j}^{\dagger}e^{i\omega_{j}t}B_{j}(t)
+\mathrm{h.c.}
\end{equation}

The total density operator in the interaction picture satisfies
\begin{equation}
\frac{d}{dt}\rho_{\mathrm{tot}}^{\mathrm{I}}(t)
=
-i\left[H_{\mathrm{eff}}+H_{\mathrm{irr}}^{\mathrm{I}}(t),\rho_{\mathrm{tot}}^{\mathrm{I}}(t)\right].
\end{equation}
Under the Born-Markovian approximation, we assume
\begin{equation}
\rho_{\mathrm{tot}}^{\mathrm{I}}(t)\approx \rho_{\mathrm{S}}^{\mathrm{I}}(t)\otimes \rho_{B},
\end{equation}
where \(\rho_{\mathrm{S}}^{\mathrm{I}}(t)\) is the reduced density operator of the system and \(\rho_{B}\) is the thermal state of the bath. Tracing out the bath degrees of freedom yields
\begin{align}
\frac{d}{dt}\rho_{I}(t)
=& -i\left[H_{\mathrm{eff}},\rho_{I}(t)\right]  - \int_{0}^{\infty} d\tau\, \mathrm{Tr}_{B}
  \left[
    H_{\mathrm{irr}}^{\mathrm{I}}(t),
  \right. \nonumber\\
& \left.
    \left[
      H_{\mathrm{irr}}^{\mathrm{I}}(t-\tau),
      \rho_{I}(t)\otimes \rho_{B}
    \right]
  \right].
\end{align}

For a thermal bath, the correlation functions are
\begin{equation}
\langle b_{j}^{\dagger}(\omega)b_{k}(\omega')\rangle
=
\delta_{jk}\delta(\omega-\omega')\,\bar{n}_{j}(\omega),
\end{equation}
and
\begin{equation}
\langle b_{j}(\omega)b_{k}^{\dagger}(\omega')\rangle
=
\delta_{jk}\delta(\omega-\omega')
\left[\bar{n}_{j}(\omega)+1\right],
\end{equation}
where
\begin{equation}
\bar{n}_{j}(\omega)=\frac{1}{e^{\beta_{j}\omega}-1},
\end{equation}
and $\beta = 1/(k_B T)$, with $k_B$ the Boltzmann constant and $T$ the temperature of the thermal bath.
Within the Markovian approximation, the dissipative rates are evaluated at the corresponding system frequencies, giving the downhill and uphill rates respectively
\begin{align}
\Gamma_{j}^{-}&=\gamma_{j}\left(\bar{n}_{j}+1\right),\\
\Gamma_{j}^{+}&=\gamma_{j}\bar{n}_{j},
\end{align}
with
\begin{equation}
\bar{n}_{j}= \bar{n}_{j}(\omega_{j})
=
\frac{1}{e^{\beta_{j}\omega_{j}}-1}.\quad 
\end{equation}

Neglecting the Lamb shift, the quantum master equation of the system takes the Lindblad-form as
\begin{align}
\frac{d}{dt}\rho
=& -i\left[H_{\mathrm{eff}},\rho\right]
+\sum_{j=s,p}\gamma_{j}\left(\bar{n}_{j}+1\right)\mathcal{D}[a_{j}]\rho \nonumber\\
& +\sum_{j=s,p}\gamma_{j}\bar{n}_{j}\mathcal{D}[a_{j}^{\dagger}]\rho,
\label{eq:finiteT-master}
\end{align}
where
\begin{equation}
\mathcal{D}[L]\rho
=
L\rho L^{\dagger}
-\frac{1}{2}L^{\dagger}L\rho
-\frac{1}{2}\rho L^{\dagger}L.
\end{equation}
The terms proportional to \(\bar{n}_{j}\) describe thermal excitation from the environment, while the terms proportional to \(\bar{n}_{j}+1\) correspond to decay processes including both spontaneous and stimulated emission.

In a typical DOPO, the signal wavelength is $\lambda=1550~\mathrm{nm}$, corresponding to an energy $E_s = hc/\lambda \approx 1.28\times10^{-19}~\mathrm{J}$. At the room temperature, the thermal energy is $k_B T \approx 4.14\times10^{-21}~\mathrm{J}$. Therefore, the thermal occupation number is
\begin{equation}
\bar{n}_s = \frac{1}{e^{E_s/k_B T}-1} \approx 3.5\times10^{-14} \ll 1.
\end{equation}
A similar estimation holds for the pump mode. Consequently, for typical DOPO parameters at the room temperature, the thermal-excitation terms proportional to $\bar{n}_s$ and $\bar{n}_p$ in Eq.~(\ref{eq:finiteT-master}) are negligibly small and the finite-temperature master equation effectively reduces to its zero-temperature form.

Therefore, under typical DOPO operating conditions, Eq.~(\ref{eq:finiteT-master}) is well approximated by the zero-temperature master equation used in the main text as
\begin{align}
\frac{d\rho}{dt}
&= -i\bigl[H_{\mathrm{eff}},\rho\bigr]
+\gamma_s\Bigl(a_s\,\rho\,a_s^\dagger
-\tfrac{1}{2}\{a_s^\dagger a_s,\rho\}\Bigr) \nonumber\\
&\quad +\gamma_p\Bigl(a_p\,\rho\,a_p^\dagger
-\tfrac{1}{2}\{a_p^\dagger a_p,\rho\}\Bigr).
\end{align}

\section{Determination of the Pump Threshold}
\label{sec:AppendixB}
In a CIM, the DOPO has a pump threshold. Below the threshold, the signal field mainly exhibits the characteristics of quantum squeezing. Above the threshold, parametric oscillation occurs, and the energy of the optical field is further enhanced. For a QB, our goal is to store and extract energy, which requires the DOPO to operate above the pump threshold. Meanwhile, the pump intensity should not be much higher than the pump threshold, as this would damage the optical components.

From Eq.~(\ref{eq:Hamiltonian}), by transforming to the rotating frame with respect to
\begin{equation}
U(t) = \exp\!\left[i \omega_{p} t \left({a}_{p}^{\dagger}{a}_{p} + \tfrac{1}{2}{a}_{s}^{\dagger}{a}_{s}\right)\right],
\end{equation}
we obtain the Heisenberg--Langevin equations as \cite{breuer2002,Wang2013PRA}
\begin{align}
\frac{d}{dt}\tilde{a}_{s}&=-\frac{\gamma_{s}}{2}\tilde{a}_{s}+\kappa\tilde{a}_{s}^{\dagger}\tilde{a}_{p}+\sqrt{\gamma_{s}}\tilde{B}_{s}, \label{eq:10}\\
\frac{d}{dt}\tilde{a}_{p}&=-\frac{\gamma_{p}}{2}\tilde{a}_{p}-\frac{\kappa}{2}\tilde{a}_{s}^{2}+\sqrt{\gamma_{p}}F_{p}+\sqrt{\gamma_{p}}\tilde{B}_{p},\label{eq:11}
\end{align}
where $\tilde{a}_{j} = {a}_{j}\exp(i\omega_{j}t)$ and $\tilde{B}_{j} ={B}_{j}\exp(i\omega_{j}t) \, (j = s,p)$ are the slowly-varying operators. 
In Eq.~(\ref{eq:10}), the first term on the right hand side represents the cavity loss, and the second term corresponds to the nonlinear $\chi^{(2)}$ process that converts the pump photons into the signal photons, and the third term represents the vacuum fluctuation of the signal field. In Eq.~(\ref{eq:11}), the first term of the right hand side also denotes the cavity loss, and the second term corresponds to the nonlinear conversion of the pump photons into the signal photons, and the third term represents the externally-injected pump drive, and the last term accounts for the vacuum fluctuation of the pump field.

By taking the expectation value and noting that $\langle \tilde{B}_{j} \rangle = 0$, we can obtain
\begin{align}
\ensuremath{\dot{\alpha}_{s}}&=-\frac{\gamma_{s}}{2}\alpha_{s}+\kappa\alpha_{s}^{*}\alpha_{p}, \\
\ensuremath{\dot{\alpha}_{p}}&=-\frac{\gamma_{p}}{2}\alpha_{p}-\frac{\kappa}{2}\alpha_{s}^{2}+\sqrt{\gamma_{p}}F_{p},
\end{align}
where $\alpha_{j} = \langle \tilde{a}_{j} \rangle$ represents the mean amplitude of each intracavity mode. 
When the system reaches a steady state, i.e., $\dot{\alpha}_{j} = 0 \, (j = s,p),$ and $\alpha_{s}(\infty) = 0,$ it indicates that there is no coherent component in the signal field. 
In this case, $\alpha_{p}(\infty) = 2F_{p}/\sqrt{\gamma_{p}}.$ This leads to
\begin{equation}
\frac{d}{dt}\left(\begin{array}{c}
\alpha_{s}\\
\alpha_{s}^{*}
\end{array}\right)=\begin{pmatrix}-\frac{\gamma_{s}}{2} & \kappa\alpha_{p}\\
\kappa\alpha_{p}^{*} & -\frac{\gamma_{s}}{2}
\end{pmatrix}\left(\begin{array}{c}
\alpha_{s}\\
\alpha_{s}^{*}
\end{array}\right),
\end{equation}
whose eigenvalues are $\lambda_{\pm}=-\gamma_{s}/2\pm\kappa|\alpha_{p}(\infty)|.$ 
The general solution can be written as $\alpha_{s}(t) = \sum_{j=\pm}c_{j}\exp(\lambda_{j}t)$. 
To ensure that $\alpha_{s}(\infty)=0$, the condition $\lambda_{+} = -\gamma_{s}/2 + \kappa|\alpha_{p}(\infty)| < 0$ must be satisfied. 
The pump threshold is thus 
$F_{p}^{(\mathrm{th})}=\gamma_{s}\sqrt{\gamma_{p}}/4\kappa$ \cite{walls2008}. 
For $F_{p} > F_{p}^{(\mathrm{th})},$ a coherent component appears in the signal field.

\section{Finite-dimensional Truncation of the Photon-Number Space}
\label{sec:AppendixC}

\begin{figure}[htp]
\includegraphics[width=8.5cm]{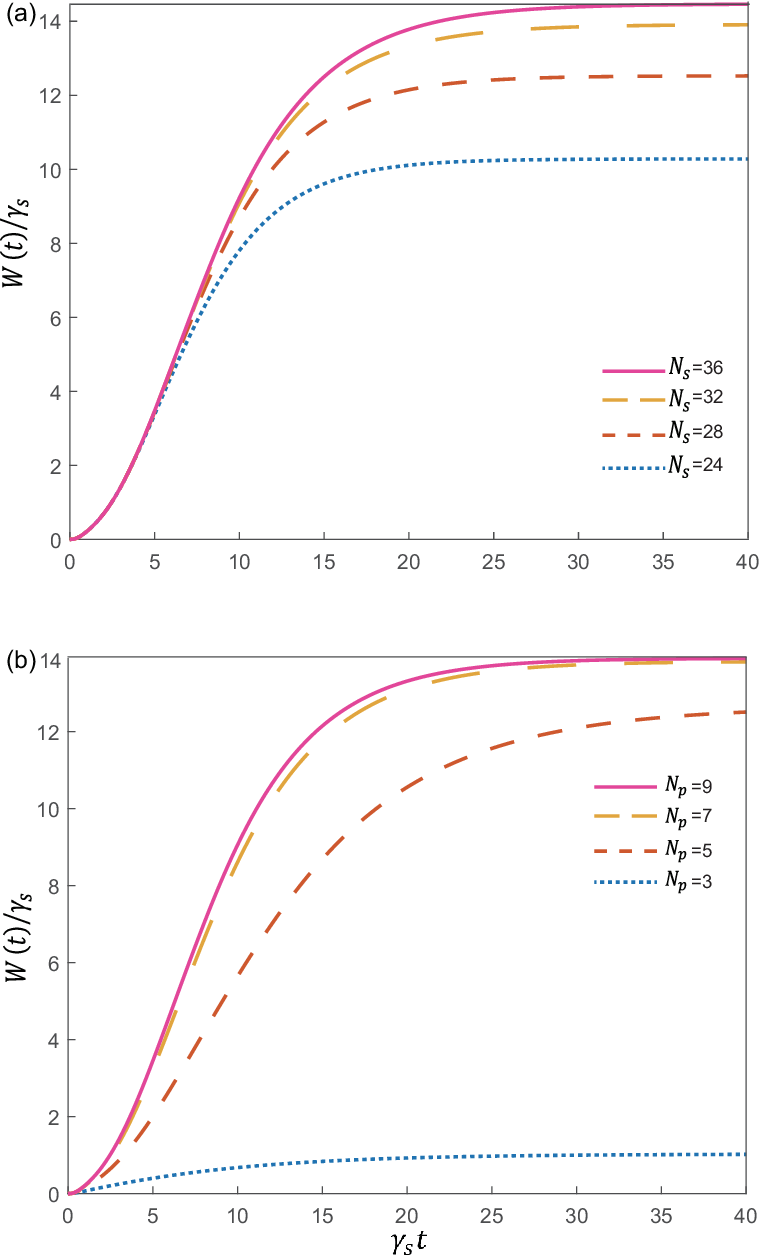}
\caption{(a) The ergotropy of the QB for different truncations of the signal field. The blue dotted, red short-dashed, yellow long-dashed, pink solid lines correspond to $N_{s}=24,\,28,\,32,\,36$ and $N_{p}=9$, respectively. (b) The ergotropy of the QB for different truncations of the pump field. The blue dotted, red short dashed, yellow long-dashed, pink solid lines correspond to $N_{p}=3,\,5,\,7,\,9$ and $N_{s}=32$, respectively. }\label{fig:10}
\end{figure}

By substituting the initial state $\rho(0)=|0\rangle_{ss}\!\langle 0|\otimes |0\rangle_{pp}\!\langle 0|$ into  Eq.~(\ref{eq:master}), and taking the partial trace over the pump field, we can obtain $\rho_{s}(t)$. Substituting it into Eq.~(\ref{eq:5}), the time evolution of the ergotropy can be obtained. In Fig.~\ref{fig:10}, we set $N_{p}=9$ and vary $N_{s}$, which denote the truncation dimensions of the photon-number space for the pump and signal fields, respectively. As $N_{s}$ increases, the value of the ergotropy gradually approaches 14.5. Similarly, we fix $N_{s}=32$ and take different values of $N_{p}$. We also find that the ergotropy increases with $N_{p}$ and gradually approaches 14. Moreover, due to the existence of driving and dissipation in the system, the ergotropy eventually becomes stable. In the following numerical calculations, the truncation is chosen as $N_{p}=9$ and $N_{s}=32$.

\section{Decay of Coherent and Incoherent Ergotropy in a TLS}
\label{sec:AppendixD}

For clarity, in this Appendix we illustrate the definitions of coherent and incoherent ergotropy by using a TLS as a QB. To keep the notation consistent with the main text, $\rho_{\mathrm{TLS}}(t)$ denotes the density matrix of the TLS, and $\varrho_{\mathrm{TLS}}(t)$ is the dephased state obtained from $\rho_{\mathrm{TLS}}(t)$ by removing all off-diagonal elements in the energy eigenbasis, and $\tilde{\rho}_{\mathrm{TLS}}(t)$ is the passive state constructed from $\rho_{\mathrm{TLS}}(t)$, and $\tilde{\varrho}_{\mathrm{TLS}}(t)$ is the passive state constructed from $\varrho_{\mathrm{TLS}}(t)$. The incoherent and coherent parts of ergotropy are then defined as
\begin{align}
W_{\mathrm{TLS}}^{i}(t) &= \mathrm{Tr}\!\left[{H}_{\mathrm{TLS}}\rho_{\mathrm{TLS}}(t)\right]
\!\!-\!\!\mathrm{Tr}\!\left[{H}_{\mathrm{TLS}}\tilde{\varrho}_{\mathrm{TLS}}(t)\right], \\
W_{\mathrm{TLS}}^{c}(t) &= \mathrm{Tr}\!\left[{H}_{\mathrm{TLS}}\tilde{\varrho}_{\mathrm{TLS}}(t)\right]
\!\!-\!\!\mathrm{Tr}\!\left[{H}_{\mathrm{TLS}}\tilde{\rho}_{\mathrm{TLS}}(t)\right].
\end{align}

As an illustrative example, we consider a TLS with excited state $|e\rangle$ and ground state $|g\rangle$. Its Hamiltonian is
\begin{align}
{H}_{\mathrm{TLS}} =
\begin{pmatrix}
0 & 0\\
0 & \omega
\end{pmatrix},
\end{align}
where $\omega$ is the level spacing. The density matrix of the TLS can be written as
\begin{align}
\rho_{\mathrm{TLS}}(t)=
\begin{pmatrix}
1-p(t) & c(t)\\
c^{*}(t) & p(t)
\end{pmatrix},\label{eq:rho_TLS}
\end{align}
where $p(t)$ is the excited-state population and $c(t)$ is the off-diagonal coherence term.

By diagonalizing $\rho_{\mathrm{TLS}}(t)$, we obtain its eigenvalues
\begin{align}
\lambda_{\pm}(t)=\frac{1\pm\sqrt{\left(2p(t)-1\right)^2+4|c(t)|^2}}{2},
\end{align}
and hence the passive state
\begin{align}
\tilde{\rho}_{\mathrm{TLS}}(t)=
\begin{pmatrix}
\lambda_{+}(t) & 0\\
0 & \lambda_{-}(t)
\end{pmatrix}.
\end{align}
The dephased state is
\begin{align}
\varrho_{\mathrm{TLS}}(t)=
\begin{pmatrix}
1-p(t) & 0\\
0 & p(t)
\end{pmatrix}.
\end{align}

The stored energy of the TLS is
\begin{align}
E_{\mathrm{TLS}}(t)
=\mathrm{Tr}\!\left[{H}_{\mathrm{TLS}}\rho_{\mathrm{TLS}}(t)\right]
=\omega p(t),
\end{align}
while the passive energy associated with $\tilde{\rho}_{\mathrm{TLS}}(t)$ is
\begin{align}
\mathrm{Tr}\!\left[{H}_{\mathrm{TLS}}\tilde{\rho}_{\mathrm{TLS}}(t)\right]
=\omega \lambda_{-}(t).
\end{align}

We now discuss two regimes.
When $p(t)\le 1/2$, the ground-state population is dominant, the dephased state is already passive, i.e.,
\begin{align}
\tilde{\varrho}_{\mathrm{TLS}}(t)=
\begin{pmatrix}
1-p(t) & 0\\
0 & p(t)
\end{pmatrix}.
\end{align}
Therefore,
\begin{align}
\mathrm{Tr}\!\left[{H}_{\mathrm{TLS}}\tilde{\varrho}_{\mathrm{TLS}}(t)\right]
&=\omega p(t),\\
W_{\mathrm{TLS}}^{i}(t)&=0.
\end{align}
In this regime, no incoherent ergotropy can be extracted from the population inversion, and the total ergotropy is entirely given by the coherent part,
\begin{align}
W_{\mathrm{TLS}}(t)=W_{\mathrm{TLS}}^{c}(t)
=\omega p(t)-\omega \lambda_{-}(t).
\end{align}

When $p(t)\ge 1/2$, the populations in the dephased state must be rearranged to form the passive state, i.e.,
\begin{align}
\tilde{\varrho}_{\mathrm{TLS}}(t)=
\begin{pmatrix}
p(t) & 0\\
0 & 1-p(t)
\end{pmatrix}.
\end{align}
In this case, the incoherent part becomes
\begin{align}
W_{\mathrm{TLS}}^{i}(t)
&=\omega p(t)-\omega\bigl(1-p(t)\bigr)
=\omega\bigl(2p(t)-1\bigr),
\end{align}
while the coherent part is
\begin{align}
W_{\mathrm{TLS}}^{c}(t)
=\omega\bigl(1-p(t)\bigr)-\omega\lambda_{-}(t).
\end{align}

In particular, when the coherence vanishes, i.e., $c(t)=0$, one has
\begin{align}
\rho_{\mathrm{TLS}}(t)&=\varrho_{\mathrm{TLS}}(t), \\
\tilde{\rho}_{\mathrm{TLS}}(t)&=\tilde{\varrho}_{\mathrm{TLS}}(t),
\end{align}
and therefore
\begin{align}
\mathrm{Tr}\left[{H}_{\mathrm{TLS}}\tilde{\varrho}_{\mathrm{TLS}}(t)\right]
&=\mathrm{Tr}\left[{H}_{\mathrm{TLS}}\tilde{\rho}_{\mathrm{TLS}}(t)\right],\\
W_{\mathrm{TLS}}^{c}(t)&=0.
\end{align}
This explicitly shows that the coherent part of ergotropy originates from the off-diagonal coherence term $c(t)$.

Next, we examine the effect of dissipation on both coherent and incoherent ergotropy. Under amplitude damping, the Lindblad-form master equation reads \cite{breuer2002}
\begin{align}
\frac{d\rho_{\mathrm{TLS}}(t)}{dt}
=&-i\left[{H}_{\mathrm{TLS}},\rho_{\mathrm{TLS}}(t)\right]+\gamma_{s}\left(
\sigma^{-}\rho_{\mathrm{TLS}}(t)\sigma^{+} \right.\nonumber\\
& \left.
-\frac{1}{2}\left\{\sigma^{+}\sigma^{-},\rho_{\mathrm{TLS}}(t)\right\}
\right),
\end{align}
where $\gamma_s$ is the relaxation rate. Substituting Eq.~(\ref{eq:rho_TLS}) into the above equation, we obtain
\begin{align}
\dot{p}(t)&=-\gamma_s p(t),\\
\dot{c}(t)&=\left(i\omega-\frac{\gamma_s}{2}\right)c(t).
\end{align}
The corresponding solutions are
\begin{align}
p(t)&=p(0)e^{-\gamma_s t},\\
c(t)&=c(0)e^{\left(i\omega-\frac{\gamma_s}{2}\right)t},
\end{align}
which yields
\begin{align}
|c(t)|^2=|c(0)|^2 e^{-\gamma_s t}.
\end{align}

For $p(t)\ge 1/2$, the incoherent part becomes
\begin{align}
W_{\mathrm{TLS}}^{i}(t)
=\omega\bigl(2p(t)-1\bigr)
=\omega\left(2p(0)e^{-\gamma_s t}-1\right).
\end{align}
Meanwhile, the coherent part can be written as
\begin{align}
W_{\mathrm{TLS}}^{c}(t)
=&\,\frac{\omega}{2}\left[
\sqrt{\left(2p(0)e^{-\gamma_s t}-1\right)^2+4|c(0)|^2e^{-\gamma_s t}}
\right.\nonumber\\
&\left.
-\left|2p(0)e^{-\gamma_s t}-1\right|
\right].
\end{align}
These expressions provide a simple explanation for the different behaviors of the two contributions. Under amplitude damping, the incoherent part $W_{\mathrm{TLS}}^{i}(t)$ is fully determined by the population inversion $2p(t)-1$, and therefore decreases rapidly as the excited-state population decays, vanishing once $p(t)\le 1/2$. By contrast, the coherent part $W_{\mathrm{TLS}}^{c}(t)$ remains nonzero as long as the off-diagonal coherence survives, i.e., $c(t)\neq 0$. Therefore, within this amplitude-damping model, the coherent contribution generally exhibits stronger robustness against dissipation than the incoherent one.
%
%






%

\end{document}